\pdfoutput=1

\documentclass[3p,twoside,twocolumn,number,sort&compress,final]{elsarticle}

\usepackage{amsmath,amssymb,stix,bm}
\usepackage[colorlinks]{hyperref} 

\makeatletter
\renewcommand\@makefntext[1]{\leftskip=0.0em\hskip-0.5em\@makefnmark{#1}}
\def\ps@pprintTitle{
   \let\@oddhead\@empty
   \let\@evenhead\@empty
   \def\@oddfoot{\footnotesize
       Submitted to Elsevier on January 15, 2019 \hfill Published in \href{https://doi.org/10.1016/j.physleta.2019.04.033}{Physics Letters A {383}~(19) (2019) 2264--2266}}
   \let\@evenfoot\@oddfoot}
\makeatother

\newcommand{\maj}{\mathrm{maj}\mkern1mu}
\newcommand{\gkl}{\mathrm{GKL}\mkern1mu}

\let\pagenbr\thepage
\renewcommand*\thepage{\small\pagenbr}
\begin{document} \small

\begin{frontmatter}

\title{Simply modified GKL density classifiers that reach consensus faster}

\author[lptms,each]{J. Ricardo G. Mendon\c{c}a\href{https://orcid.org/0000-0002-5516-0568}{\includegraphics[viewport=-2 0 36 32, scale=0.30]{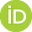}}}\corref{permanent}

\address[lptms]{LPTMS, UMR 8626, CNRS, Universit\'{e} Paris-Sud, Universit\'{e} Paris Saclay, 91405 Orsay CEDEX, France}

\address[each]{Escola de Artes, Ci\^{e}ncias e Humanidades, Universidade de S\~{a}o Paulo, Rua Arlindo Bettio 1000, 03828-000 S\~{a}o Paulo, SP, Brazil \vspace{-12pt}}

\cortext[permanent]{Permanent address: Escola de Artes, Ci\^{e}ncias e Humanidades, \linebreak Universidade de S\~{a}o Paulo, SP, Brazil. Email: \href{mailto:jricardo@usp.br}{\texttt{jricardo@usp.br}}.}

\begin{abstract}
The two-state Gacs-Kurdyumov-Levin (GKL) cellular automaton has been a staple model in the study of complex systems due to its ability to classify binary arrays of symbols according to their initial density. We show that a class of modified GKL models over extended neighborhoods, but still involving only three cells at a time, achieves comparable density classification performance but in some cases reach consensus more than twice as fast. Our results suggest the time to consensus (relative to the length of the CA) as a complementary measure of density classification performance.
\end{abstract}


\end{frontmatter}


In 1978, Gacs, Kurdyumov, and Levin (GKL) introduced the density classification problem for cellular automata (CA) in the literature \cite{gkl,russkiye,maes}. The problem consists in classifying arrays of symbols according to their initial density using local rules, and is completed successfully if all the cells of the CA converge to the initial majority state in linear time in the size of the input array. Density classification is a nontrivial task for CA composed of autonomous and memoryless cells because the cells have to achieve a global consensus cooperating locally; emergence of collective behavior is required. The GKL two-state model, or GKL-II for short, became a staple model in the theory of complex systems related with the concepts of communication, efficiency, and emergence \cite{mitchell,crutch,mechanics}. It has been demonstrated that the density classification problem cannot be solved correctly $100\%$ of the times by uniform two-state CA, although no upper bound on the maximum possible efficiency has been set \cite{landbelew,busic}. Solutions involving nonuniform CA and less strict criteria for what a solution to the problem means exist \cite{fuks,sipcapron}. Recent reviews on the density classification problem for CA are given in \cite{ppreview,fates}.


The GKL-II CA is a finite one-dimensional array of $n \geq 4$ cells under periodic boundary conditions evolving by the action of a transition function $\Phi_{\mathrm{II}} \colon \{0,1\}^{n} \to \{0,1\}^{n}$ that given the state $\bm{x}^{t} = (x_{1}^{t}, \ldots, x_{n}^{t})$ of the CA at instant $t$ determines its state $\bm{x}^{t+1} = \Phi_{\mathrm{II}}(\bm{x}^{t})$ at instant $t+1$ by the majority rule
\begin{equation}
\label{eq:gklii}
x_{i}^{t+1} = 
   \begin{cases}
      \maj(x_{i-3}^{t}, x_{i-1}^{t}, x_{i}^{t}), & \text{if } x_{i}^{t}=0, \\
      \maj(x_{i}^{t}, x_{i+1}^{t}, x_{i+3}^{t}), & \text{if } x_{i}^{t}=1,
   \end{cases}
\end{equation}
where $\maj(p,q,r)=\lfloor\frac{1}{2}(p+q+r)\rfloor$ for $0$-$1$ variables $p$, $q$, $r$. The CA classifies density if $\bm{x}^{t} \to \bm{0}=(0, \dots, 0)$ or $\bm{1} = (1, \dots, 1)$ depending whether, respectively, the initial density $\rho^{0} = n^{-1}\sum_{i}x_{i}^{0} < 1/2$ or $\rho^{0} > 1/2$. We do not require a definite behavior when $\rho^{0} = 1/2$. The CA is supposed to reach consensus in $O(n)$ time steps. In \cite{gkl,maes}, the authors prove that the GKL-II CA on the infinite lattice $\mathbb{Z}$ displays the eroder property, washing out finite islands of the minority phase in finite time and eventually leading the CA to one of the two invariant states $\bm{0}$ or $\bm{1}$. In an array of $n=149$ cells (odd length to avoid ties), GKL-II scores an average density classification performance of $81.5\%$ over random initial conditions with each cell initialized in the state $0$ or $1$ equally at random (Bernoulli product measure), taking on average $86 \sim 0.576\,n $ time steps to reach consensus. Details on the GKL-II performance are given in \cite{mitchell,crutch,mechanics,ppreview,fates,assembly,gkliv}.


\begin{table*}[t] \small
\setlength{\tabcolsep}{5pt}
\caption{\label{tab:gkljk}Best density classification performances of $\gkl(j,k)$ in the range $1 \leq j \leq 5$, $j < k \leq 15$ in an array of $n=299$ cells averaged over $10^{6}$ random initial configurations near the critical density $\rho^{0}=1/2$. The uncertainty in the performance $\langle f \rangle$ is $\pm 0.0004$. GKL-II figures are displayed in bold for comparison.}
\centering
\begin{tabular}{ccccccccccccc}
\hline
$(j,k)$ & $(4,12)$ & $(3,9)$ & $(2,6)$ & $(5,15)$ & $\mathbf{(1,3)}$ & $(1,9)$ & $(1,11)$ & $(2,14)$ & $(2,10)$ & $(3,15)$ & $(1,7)$ & $(1,5)$ \\
\hline
$\langle f \rangle$ & $0.7926$ & $0.7922$ & $0.7921$ & $0.7920$ & $\mathbf{0.7917}$ & $0.7893$ & $0.7875$ & $0.7874$ & $0.7873$ & $0.7869$ & $0.7868$ & $0.7865$ \\
$\langle t^{*} \rangle/n$ & $0.5843$ & $0.5848$ & $0.5844$ & $0.5849$ & $\mathbf{0.5848}$ & $0.2633$ & $0.2289$ & $0.3270$ & $0.4123$ & $0.4123$ & $0.3269$ & $0.4122$ \\
\hline
\end{tabular}
\end{table*}

We now modify the neighborhood in the GKL-II. Instead of evaluating the majority vote of cell $i$ with its nearest $i \pm 1$ and third $i \pm 3$ neighbours, we pick neighbors $i \pm j$ and $i \pm k$, with $k > j \geq 1$. The rules for the modified CA read
\begin{equation}
\label{gkljk}
x_{i}^{t+1} =
   \begin{cases}
      \maj(x_{i-k}^{t}, x_{i-j}^{t}, x_{i}^{t}) & \text{if } x_{i}^{t}=0, \\
      \maj(x_{i}^{t}, x_{i+j}^{t}, x_{i+k}^{t}) & \text{if } x_{i}^{t}=1.
   \end{cases}
\end{equation}
We refer to this CA as $\gkl(j,k)$; $\gkl(1,3)$ recovers the original GKL-II model. To the best of our knowledge these models have never been considered in the literature before. We measured the average density classification performance $\langle f \rangle$ of $\gkl(j,k)$ over $10^{6}$ random initial states close to the critical density ($x_{i}^{0}=0$ or $1$ equally at random) in an array of $n=299$ cells to minimize finite-size effects that show up in the rules with larger $(j,k)$. Our results appear in Table~\ref{tab:gkljk}. We see that the $\gkl(j,k)$ with $k=3j$, i.\,e., the $\gkl(j,3j)$ models, all display virtually the same density classification and relative time to consensus ($\langle t^{*}\rangle/n$) performances. Otherwise, the $\gkl(1,9)$ and $\gkl(1,11)$ models display almost the same density classification performance as GKL-II but achieve consensus in about half the time. Explicitly, $\gkl(1,11)$ is just about $0.53\%$ less efficient than GKL-II but is ${\sim}2.5$ times faster. From Table~\ref{tab:gkljk} we conclude that if quality is critical, then $\gkl(4,12)$ is the best CA in its class, while if one needs speed, then $\gkl(1,9)$ or $\gkl(1,11)$ becomes the CA of choice.

\begin{figure}[t]
\centering
\includegraphics[viewport=10 12 530 430, scale=0.29, clip]{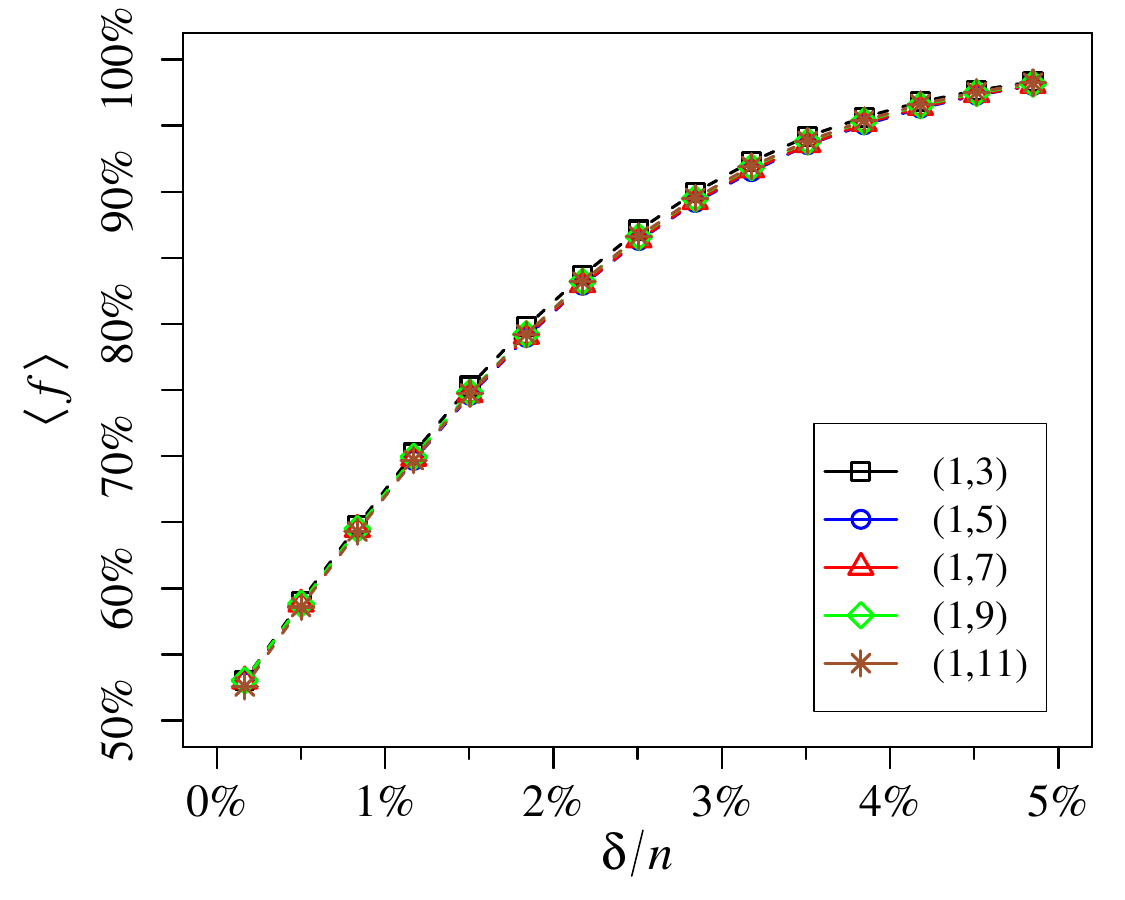} \\[3pt]
\includegraphics[viewport=10 12 530 430, scale=0.29, clip]{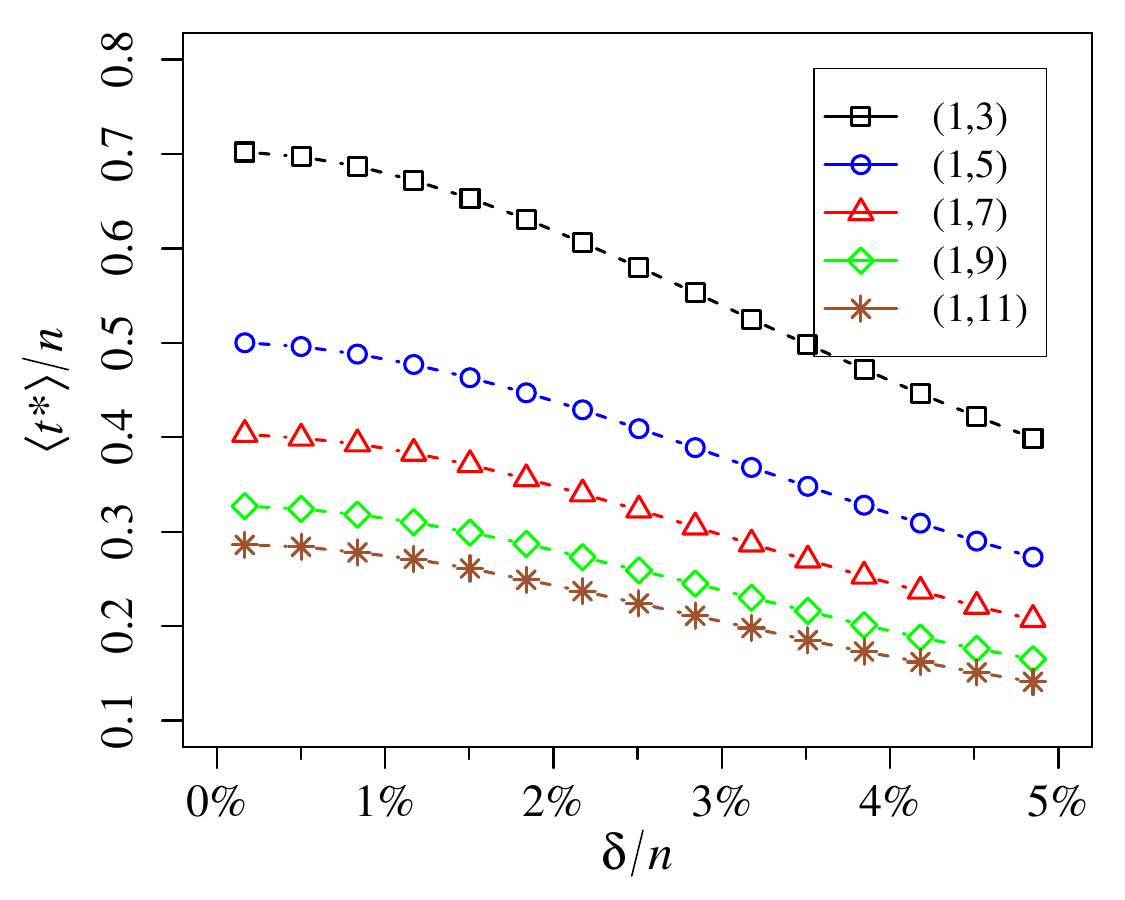}
\caption{\label{fig:imbalance}Density classification performance $\langle f \rangle$ and time to consensus $\langle t^{*} \rangle/n$ averaged over $10^{6}$ random initial configurations of some $\gkl(1,k)$ CA of length $n=299$ as a function of the relative imbalance $\delta/n=\frac{1}{2}(n_{1}-n_{0})/n$ in the initial configurations. Error bars are much smaller than the symbols shown.}
\end{figure}

\begin{figure*}[h!]
\centering
\includegraphics[viewport=30 100 480 400, scale=0.425, clip]{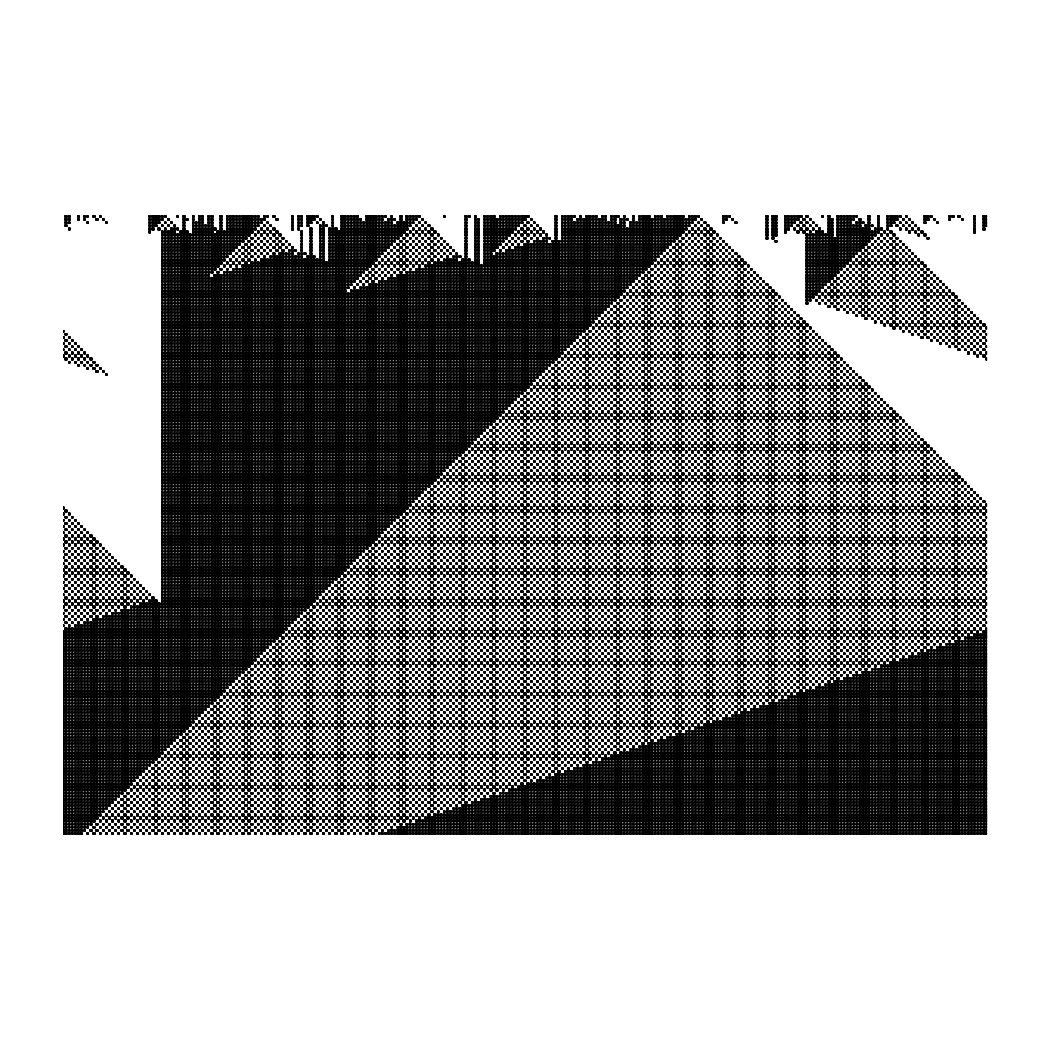}
\includegraphics[viewport=30 100 480 400, scale=0.425, clip]{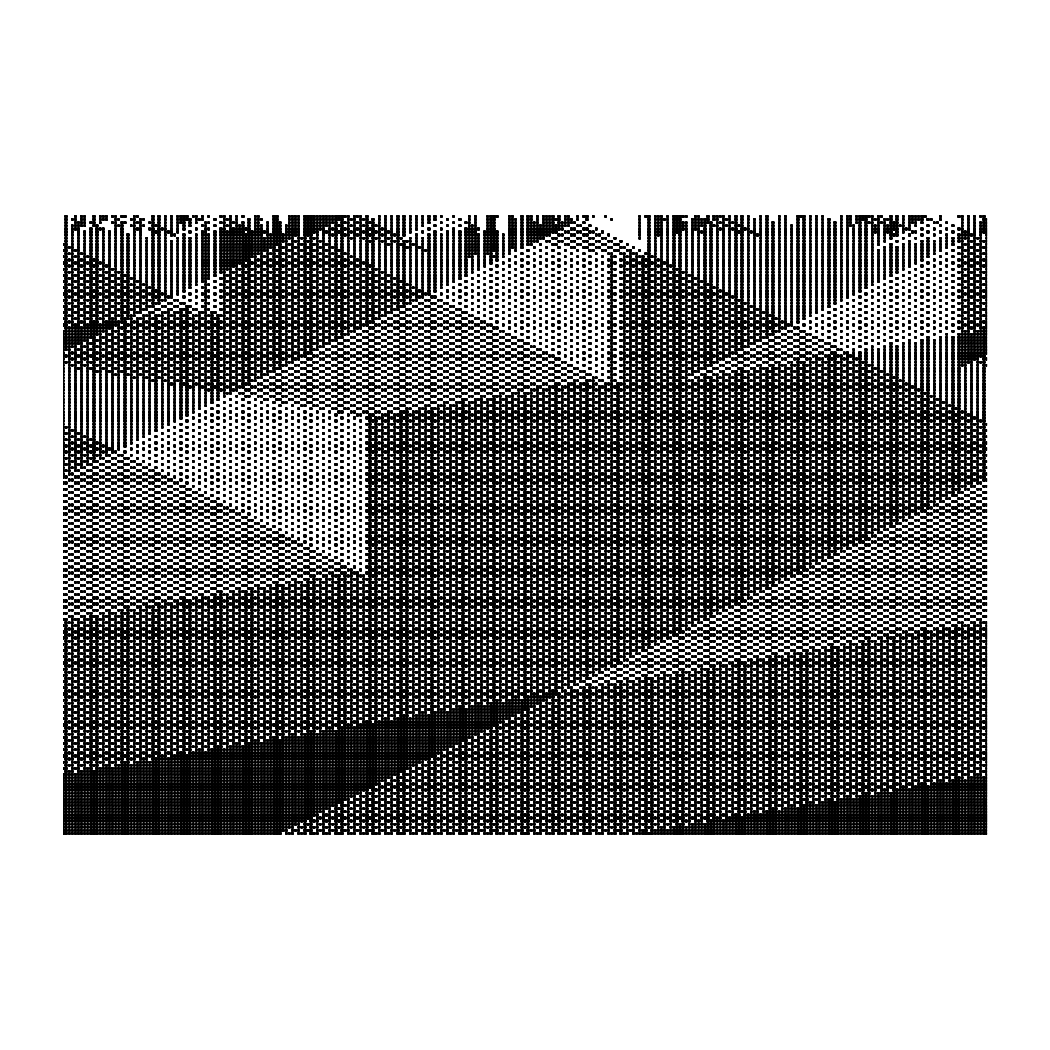}
\\[3pt]
\includegraphics[viewport=30 100 480 400, scale=0.425, clip]{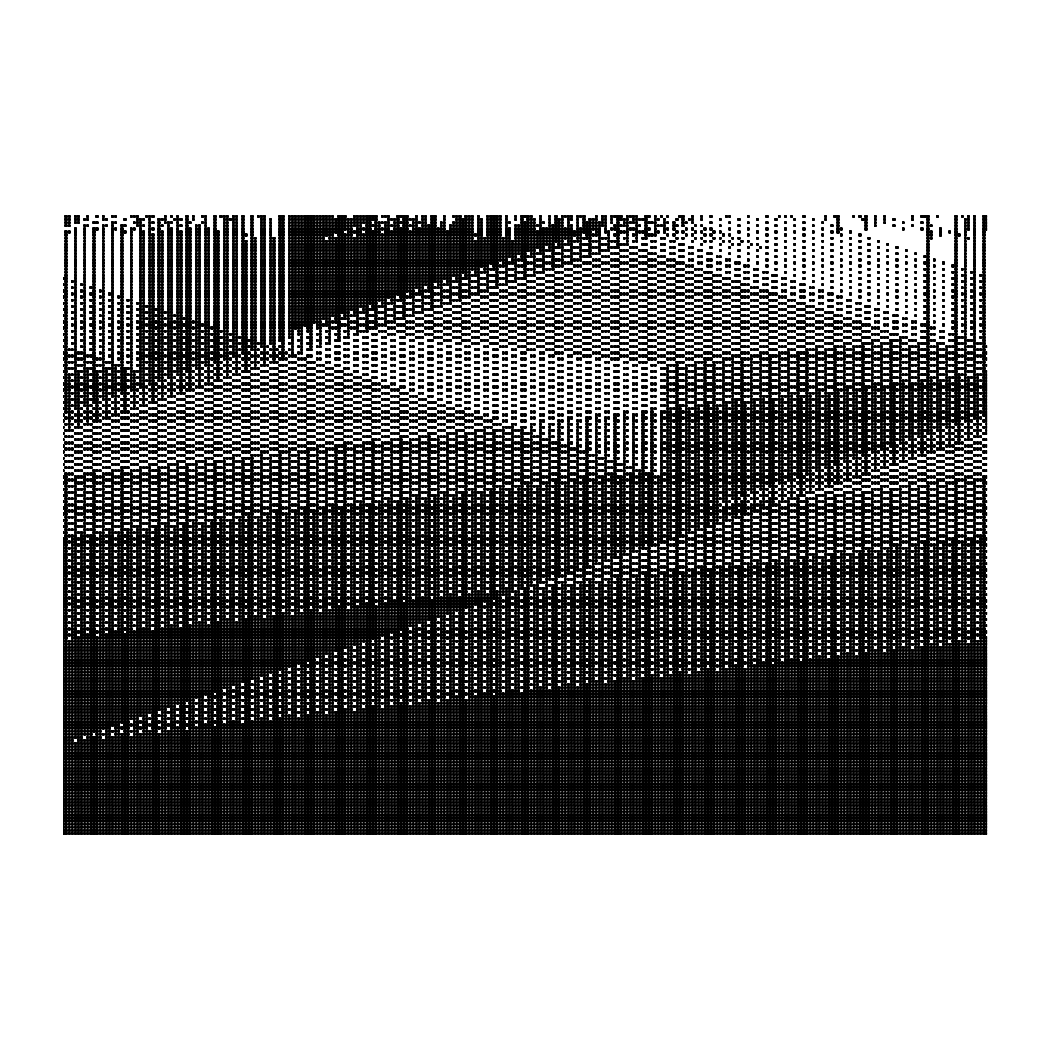}
\includegraphics[viewport=30 100 480 400, scale=0.425, clip]{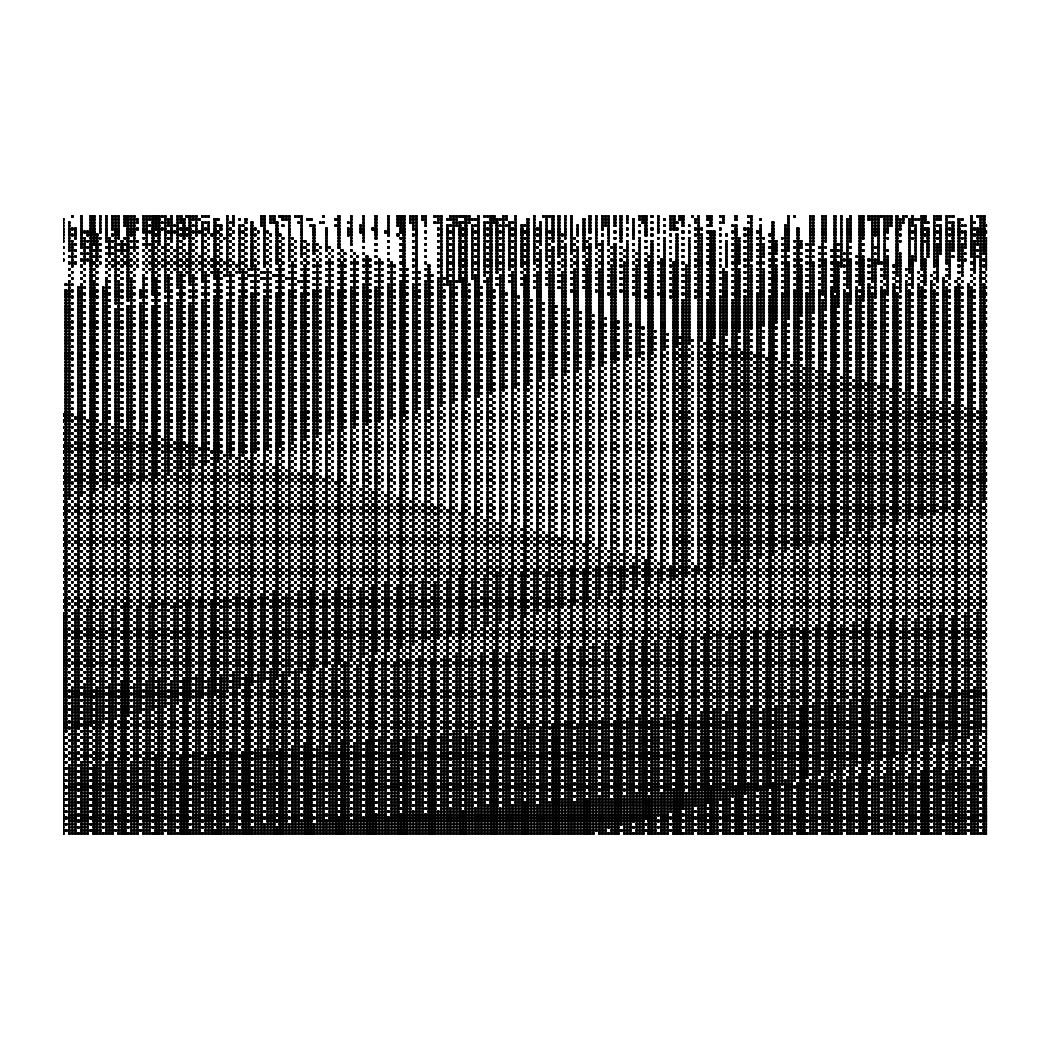}
\\[3pt]
\includegraphics[viewport=30 100 480 400, scale=0.425, clip]{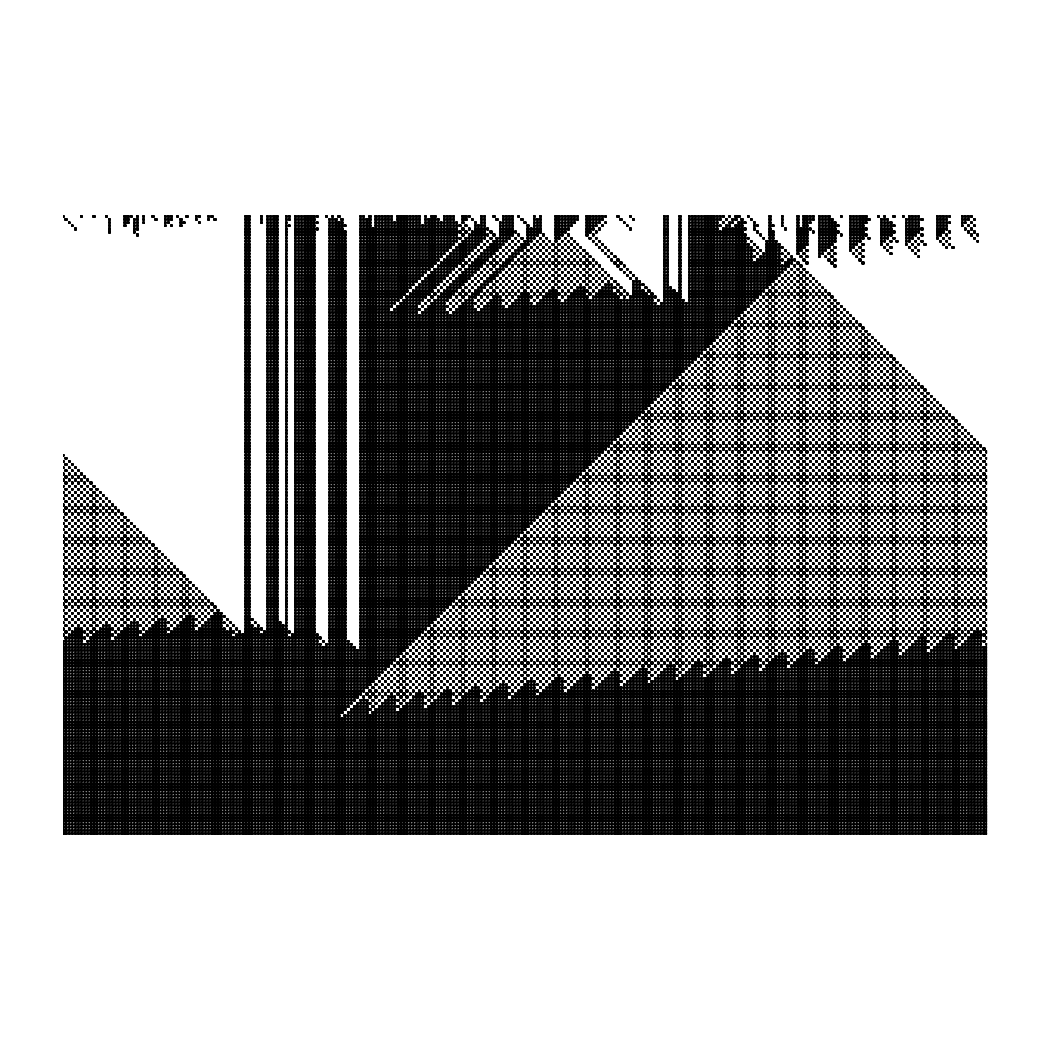}
\includegraphics[viewport=30 100 480 400, scale=0.425, clip]{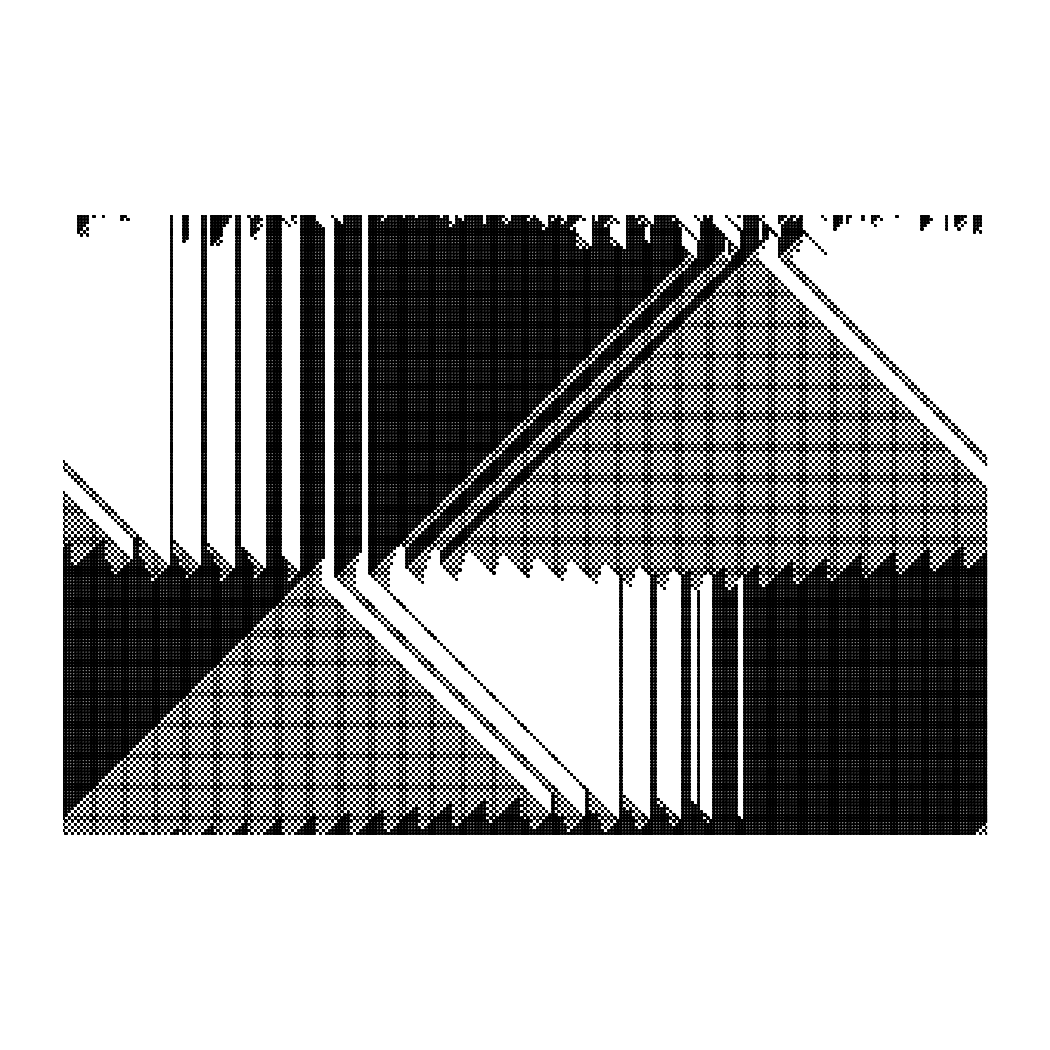}
\caption{\label{fig:patterns}Space-time diagrams of $\gkl(j,k)$ with $n=299$, $0 \leq t \leq 200$ (time flows downwards), and random initial conditions with $\rho^{0}=150/299$. From left to right, top to bottom, $(j,k)=(1,3)$ (the usual GKL-II) and $(2,6)$, $(3,9)$ and $(4,12)$, $(1,9)$ and $(1,11)$. The diagrams displayed are those that reached or would eventually reach the majority state of all-$1$ cells (in black).}
\end{figure*}

Figure~\ref{fig:imbalance} displays the average classification performance of the $\gkl(1,k)$ CA as a function of the imbalance $\delta=\frac{1}{2}(n_{1}-n_{0})$ between the number of cells in states $1$ and $0$ in the initial configuration. Here the initial density $\rho^{0}=1/2+\delta/n$ is fixed but the configurations are random. By symmetry, the performance of the CA depends only on the magnitude of $\delta$, not on its sign. The data show that the density classification performance of all these CA are close over a range of initial densities, differing significantly, however, on the time to consensus. Space-time diagrams of some $\gkl(j,k)$ CA are displayed in Figure~\ref{fig:patterns}.

We do not currently have a sound explanation for the efficient combinations of $j$, $k$ found. The efficiency of the $\gkl(j,3j)$ can be related with that of $\gkl(1,3)$ in one or more sublattices, although the fast convergence of $\gkl(1,9)$ and $\gkl(1,11)$ cannot be immediately related with any sublattice dynamics. Intuitively, in the $\gkl(j,k)$ CA information about the dynamics of the interfaces between islands of $0$s and $1$s can jump over longer distances (i.\,e., move faster) with increased $k-j$. Data from Table~\ref{tab:gkljk} for the time to consensus for $\gkl(1,k)$ with $k=3$, $5$, $7$, $9$, and $11$ corroborates this idea. Note that the metric $\langle t^{*} \rangle/n$ is not unique---one could consider the alternative timings given by $\langle t^{*} \rangle/nk$, with $k$ the radius of the CA, as well as $\langle t^{*} \rangle/nz$, with $z$ the number of cells that enter the local rule ($z=3$ for all $\gkl(j,k)$). A characterization of the ``computational mechanics'' of the $\gkl(j,k)$ CA \cite{mitchell,crutch,mechanics,gacs18} may help to understand their eroder mechanism and their efficiency better. It would also be of interest to assess the robustness of the $\gkl(j,k)$ against noise and whether the ensuing probabilistic CA display an ergodic-nonergodic transition, a long-standing unsettled issue for one-dimensional density classifiers \cite{gkl,russkiye,maes,fates,assembly,gkliv,gacs18,mairesse,roberto,siamak}. These and related questions (e.\,g., how $\langle f(n,\delta) \rangle \to 1$ for any $\delta \ne 0$ as $n \nearrow \infty$, see \cite{busic}) will be the subject of forthcoming publications.


\section*{Acknowledgments}

The author thanks Nazim Fat\`{e}s (LORIA, Nancy) for useful conversations and an anonymous reviewer for valuable suggestions improving the manuscript. The author also acknowledges the LPTMS for kind hospitality during a sabbatical leave in France and FAPESP (Brazil) for partial support through grant no.~2017/22166-9.



\end{document}